# Process Control with Highly Left Censored Data

Javier Orlando Neira Rueda [1]
Polytechnic University of Valencia, Spain. MSC.
Dept. of Applied Statistics and Operational Research, and Quality
jneirue@alumni.upv.es

Andrés Carrión García [2]
Dept. of Applied Statistics and Operational Research, and Quality.
Polytechnic University of Valencia, Spain. PhD.
Camino de Vera, s/n, edificio 7A, 46022,
Valencia, Spain
acarrion@eio.upv.es

*Abstract - The need to monitor industrial processes, detecting changes in process parameters in order to promptly correct problems that may arise, generates a particular area of interest. This is particularly critical and complex when the measured value falls below the sensitivity limits of the measuring system or below detection limits, causing much of their observations are incomplete. Such observations to be called incomplete observations or left censored data. With a high level of censorship, for example greater than 70%, the application of traditional methods for monitoring processes is not appropriate. It is required to use appropriate data analysis statistical techniques, to assess the actual state of the process at any time. This paper proposes a way to estimate process parameters in such cases and presents the corresponding control chart, from an algorithm that is also presented.*

## 1  Introduction

Industrial processes demand each time better measurement performance, and in some cases this requires measuring in the limits of equipment sensitivity. When the measured quantity is very small, and its true value falls below a certain limit of detection it is said that this value is in the category of left censored data (Klein John, 2003). With these non-detectable values, the person in charge of control may be confused as to how to treat these observations using traditional statistical methods such as Shewhart Control Chart (Mason Robert L. & Keating Jerome P., 2011).

Assuming that the detection limit is equal to a constant, C, and the engineer in charge of monitoring the process knows that the measured quantity, X, is smaller than C, but without knowing its exact value, four are the alternatives usually adopted:

1. The set of values below C are taken as zero: X = 0
2. Values below C are fixed in the mid of the interval [0, C]: X = C/2.
3. The set of values below C are taken as equal to the detection threshold: X = C
4. Values under C are ignored, and substituted by other readings over C

If we analyze censored data using the first method, it will tend to "underestimate" the true value of the mean from the sample taken. Third and fourth methods will tend to "overestimate" the true value of the mean.

If we analyze the data using the second method, we see that it is an attempt to take the middle position between methods 1 and 3. The fourth method, a part of overestimating the mean value, simply ignores undetectable, and the result can be serious. (Mason Robert L. & Keating Jerome P., 2011).

As an example of this situation, we can cite the case of monitoring pollution parameters, now very commonly controlled due to environmental protection regulation, and involving measurement of some parameters whose legal limit is very low, close to what standard measurement equipment can capture (Shumway H., 2002). This requires increasing the use of statistical techniques to reliably measure or estimate in such situations (Montejo Ulín F., 2007).

## 2 Controlling left censored statistical data

As already mentioned, there are processes where the control outputs are censored, sometimes in a large percentage, and parameter estimates are significantly biased. Even in relatively simple situations, one has to rely heavily on statistical methods for large samples and asymptotic properties.

In this section, we studied how to deal with censored data, with the objective of estimating its average and standard deviation, proposing a control chart to monitor the mean and the standard deviation in that process (which contains censored data).

It is assumed that the measured quantity, T, is normally distributed with mean, μ, and standard deviation, σ. Also, it is assumed that the observations are censored by the left (the formula is similar for right censoring (Steiner S. H. & Mackay R. Jock, 2000)).

In this case, situations with left censored observations, increases in the average of the process and increases the dispersion of the data obtained are of interest.

### *2.1  Estimating the percentage of censored data*

Consider that T is the quality characteristic that we will control for changes in variability. Consider also that T can be modeled as a normal random variable with mean μ and standard deviation σ (T~N (μ, σ)). Then T will have a probability density function (PDF)

***Equation 1***

$$f(t) = \sigma^{-1}\emptyset\left[\frac{t-\mu}{\sigma}\right]$$

The cumulative distribution function (CDF) is denoted as $\Phi[(t-\mu)/\sigma]$ and a typified value $Z=[(t-\mu)/\sigma]$ where t is the observed value. (Lawless, J.F., 1982)

The probability of censure for a random variable, T, normally distributed with mean, μ and standard deviation, σ, censored by the left of C is described as:

*Equation 2*

$$Pc = P[T \leq C] = P\left(\frac{T-\mu}{\sigma}\right) = \left(Z \leq \frac{C-\mu}{\sigma}\right) = \Phi\left(\frac{C-\mu}{\sigma}\right)$$

Then, we can call $Z_C$ to the typified point value censorship C, and $\Phi(Z_C)$, it is the function of Normal Distribution Model Typified at that point C (Martinez, 1998).

*Equation 3*

$$Z_c = \frac{C-\mu}{\sigma}$$

Thus, we can write:

*Equation 4*

$$Pc = P[T \leq C] = \Phi(Z_c) = \int_{-\infty}^{Z_c} \left(\frac{1}{\sigma\sqrt{2\pi}} e^{-\frac{(Z_c)^2}{2}}\right) dZ$$

For example, for data normally distributed N (0,1) with a fixed level by the left censorship C = 1 is obtained:

*Equation 5*

$$Pc = \Phi\left(\frac{1-0}{1}\right) = \Phi(1) = 0,8413$$

Where, Pc is the censorship ratio.

Now, using the conditional expected value CEV technique (Steiner S. H. & Mackay R. Jock, 2000), you can define a control chart.

With left-censored data, the target to the graph CEV Control (Conditional Expected Value) is to detect increases in the mean and / or increases in the standard deviation of the process. In other words, the two control charts have a single control limit as discussed later. Moreover, left censored data is very difficult to detect decreases in the process mean that such changes increase the proportion of censorship. Similarly if the proportion of censorship is greater than 50%, a decrease in the dispersion process also leads to more censored observations. Subgroups with all censored observations provide little information about changes in process parameters (Steiner S. H. & Mackay R. Jock, 2000) and may additionally generate a further biased estimate.

## 2.2 *Calculating Weights CEV for censored data to the left*

The control chart proposed in this paper, is based on replacing each censored observation by a conditional expected value denoted as Wc, which we will call "Weights CEV". These weights are based on, the sample mean and standard deviation that is plotted subgroup similarly to traditional graphics $\overline{X}$ y S.

This conditional expected value or weight Wc for left censored observations is obtained as:

*Equation 6*

$$Wc = E(T|T \leq C) = \mu - \sigma \left( \frac{\phi(Zc)}{\Phi(Zc)} \right)$$

Where the term $(\phi(Z_C)/\Phi(Z_C))$ can be denoted as the role of chance $V(Z_C)$, defined as the function of chance the probability density function (PDF) and the cumulative distribution function (CDF) (Lawless, J.F., 1982):

*Equation 7*

$$V(Zc) = \left( \frac{\phi(Zc)}{\Phi(Zc)} \right) = \left( \frac{\phi\left(\frac{C-\mu}{\sigma}\right)}{\Phi\left(\frac{C-\mu}{\sigma}\right)} \right)$$

Since $\phi(Z_C)$, the Standard Normal Probability Density Function (PDF) at the point of censorship C is:

*Equation 8*

$$\phi(Zc) = \frac{1}{\sigma\sqrt{2\pi}} e^{-\frac{\left(\frac{C-\mu}{\sigma}\right)^2}{2}}$$

$$C \in R; \mu \in R; \sigma > 0$$

Therefore, the new data used to build the control chart CEV, and estimate the new parameters, are denoted as:

The Control Chart CEV monitoring the average and standard deviation of the subgroups with weights CEV ($w_i$). It will be call Control Chart CEV $\overline{X}$ for averages and Control Chart CEV S for standard deviation. (Steiner S. H. & Mackay R. Jock, 2000). **The calculation of the weights for the censored observations depends on the parameters μ and σ under control.**

*Equation 9*

$$w_i = \begin{cases} t, \text{If } t > C \\ W_c, \text{If } t \leq C \end{cases}$$

The procedure for estimating the parameters μ and σ of a process under control, you will see later in the initial implementation and process of estimation for data monitoring with left censored observations.

*The idea of using weights CEV is based on the likelihood function given by Steiner & Mackay, who in turn are based on the book Lawless, J.F., 1982.*

## 2.3 *Process of maximum likelihood*

Like the process of maximum likelihood for censored data by the right of Steiner & Mackay, the process of maximum likelihood for censored data by the left is iterative and involves replacing each censored observation with conditional expected value.

The estimation algorithm is fed from the Equation 6 and Equation 9 represented in section 2.2. Based on these weights, the mean and the standard deviation of the process are re-estimated.

The estimated mean and standard deviation are obtained by:

*Equation 10*

$$\hat{\mu}_i = \sum_{i=1}^{n} \frac{w_i}{n}$$

*Equation 11*

$$\hat{\sigma}_i = \sqrt{\frac{\sum_{i=1}^{n}(w_i - \hat{\mu}_i)^2}{r + (n-r)\lambda(Z_c)}} \qquad (Ap1)$$

Where *r* equals the number of uncensored observations, *n* the total number of data, *i* equals the number of iteration, for which:

*Equation 12*

$$\lambda(Z_c) = \frac{\phi(Z_c)}{\Phi(Z_c)}\left[\frac{\phi(Z_c)}{\Phi(Z_c)} + z\right]$$

$\lambda(Z_c)$ It is always between 0 and 1. When it is close to 1 the percentage of censorship is small and close to 0 when the proportion of censorship is great.

To calculate estimated mean and standard deviation is proposed for the CEV Model Left. The following expressions:

*Equation 13*

$$\hat{\sigma}_i = \sqrt{\frac{\sum_{i=1}^{n}(w_i - \hat{\mu}_{i-1})^2}{r + (n-r)\lambda(Zc)}} \quad (Ap2)$$

Where *r* equals the number of uncensored observations, *n* the total number of data and:

*Equation 14*

$$\lambda(Zc) = \frac{\phi\left(\frac{C-\hat{\mu}_{i-1}}{\hat{\sigma}_{i-1}}\right)}{\Phi\left(\frac{C-\hat{\mu}_{i-1}}{\hat{\sigma}_{i-1}}\right)} \left[ \frac{\phi\left(\frac{C-\hat{\mu}_{i-1}}{\hat{\sigma}_{i-1}}\right)}{\Phi\left(\frac{C-\hat{\mu}_{i-1}}{\hat{\sigma}_{i-1}}\right)} + z \right]$$

To find the maximum likelihood estimate is iteratively applied to the formula Ap2 data until estimates converge. Figure 1 show the estimation process for the proposed model.

*Figure 1 (Estimation process)*

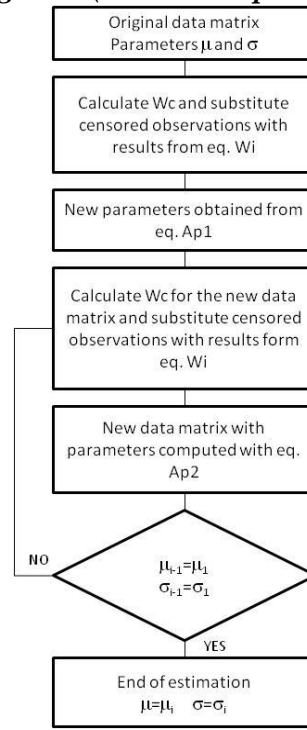

## 2.4 Control limits used for the CEV control chart, left data censored

Calculating control limits required recourse to simulation of data. Figure 2 and Figure 3 are provided for constructing graphs of the control limits of the graph CEV $\overline{X}$ and S. This limits obtained by simulating more than 1000 estimates for each level of censorship. We used a risk of false alarm of 0.0027 (type I error). (Cox & Oakes, 1984) (Mongomery, 2005)

Control limits shown on these graphs are standardized, so they give the control limit for subgroups with sample sizes (n= 3, 5, 10, 20) and Pc proportion of censorship, assuming the process is under control with mean zero and standard deviation equal to one.

Once estimated process parameters μ and σ are under control, you can place the control limits $UCL_{\overline{X}}$ and $LCL_S$, which are standardized control limits. This control limits for any issue can be obtained using the following formulas:

*Equation 15*

*Upper control limit for the chart* $CEV_{\bar{X}} = UCL_{\bar{X}}\sigma + \mu$

*Equation 16*

*Upper control limit for the chart* $CEV_S = UCL_S \sigma$

Where μ and σ are process parameters controlled.

An interpolation between the different curves allows locating a boundary of a subgroup size n; different sizes may be used. The horizontal axes for both graphs are in logarithmic scale.

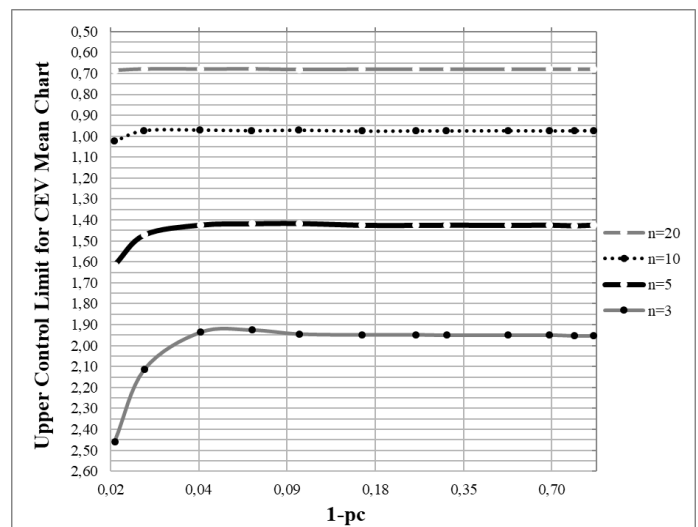

*Figure 2 (Standardized upper control limit ($UCL_{\bar{X}}$) for the graph CEV $\bar{X}$ model CEV Left)*

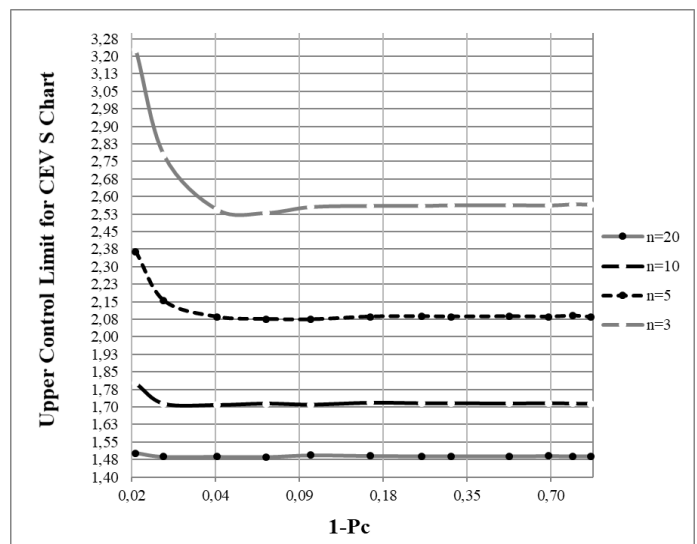

*Figure 3 (Standardized upper control limit ($UCL_S$) for de graph CEV S model CEV Left)*

Note: Values $UCL_{\overline{X}}$ y $UCL_S$ for different sample sizes and censoring proportions are detailed in Appendix.

# 3 Initial implementation, Estimation of parameters and Control chart.

Step commonly called initial implementation phase involves collecting a set of samples when the process is under control. When working with uncensored data, it is suggested to work with 100 observations or more for the initial implementation of graphics CEV $\overline{X}$ y S. This restriction ensures that the sample size estimates of the initial parameters of the process are accurate and reasonably good.

The following steps are applicable to the model CEV Left.

1. Taking K subgroups, each of size n.
2. Estimate the mean and standard deviation under control μ and σ, using the method of maximum likelihood; Figure 1
3. Determine the weight Wc CEV for censored observations with the equation given in paragraph 2.2, based on the estimation of μ and σ under control, and replace all censored observations Wc value.
4. Calculate and create control limits using the design of the *s* given for graphics CEV ($\overline{X}$ y S), plotting the averages and deviations of the subgroups.
5. Search any sign out of control in the graph (points outside the control limits). Browse process conditions, if any subgroup runaway was collected over time, repeat the procedure from step 2 if some subset out of control was removed from the sample.

*The imprecision of the estimation algorithm when censorship is high can lead to bias in the process parameters.*

Remember that in the estimation procedure process variability is calculated over full sample or matrix instead of only the dispersion within the subgroup as typically done for traditional control charts used.

As the publication (Steiner S. H. & Mackay R. Jock, 2000), maximum likelihood estimates work well for large samples. The maximum likelihood method is iterative, generating a great computational effort if the censored level is large.

## 3.1 *Example*

To demonstrate the results of the implementation, the parameter estimation and the left CEV control graph, it is presented what happens in geotextile characterization tests, specifically in flow capacity tests in the plane for the so-called drainage geocomposites; we find a case in which censorship by the left is present. This test consists in applying a confining pressure over the geotextile and evaluate the amount of water (in liters) flowing (or draining) during certain time at certain water level gradients (BAMFORTH, 2009), as shown in Figure 4

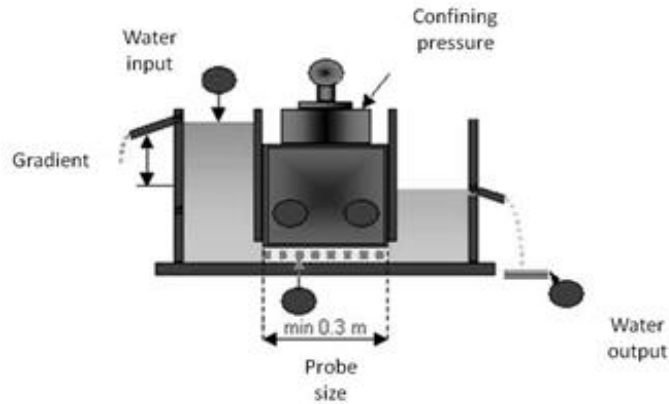

*Figure 4 (Geocomposite textile drainage test: An example of estimate with left censoring)*

The problem appears for certain combinations of the test design parameters (ambient conditions, geotextile thickness and time required for testing). For geotextile of less than 2 mm thickness and with certain water pressure gradient, the testing equipment has a limit of detection of water flow in 50ml/hour. Therefore, when one wants to monitor the performance of a geotextile whose average in-plane flow capacity is less than this limit, tests will generate left censored observations.

Consider for the process under control of a data matrix with K=100 subgroups of size n = 5 taken to estimate the mean and standard deviation under control with censorship C= 50ml/h. Table 1 shows the first 25 samples of size 5, with means and standard deviations.

The mean and the standard deviation of the process were estimated with the algorithm given in paragraph 2.3, giving the following results:

- Initial Mean Censored data:            $\mu_0 = 50,0846$
- Initial Standard Deviation Censored:   $\hat{\sigma} = 0,2720$

Applying the proposed method, estimations of mean and standard deviations are:

- Estimated Mean Under Control:              $\hat{\mu} = 49,0279$
- Estimated Standard Deviation Under Control: $\hat{\sigma} = 0,9915$

Once the process is under control and the parameters were estimated, the CEV weight is calculated with Equation 6:

$$Wc = \mu - \sigma \left( \frac{\phi\left(\frac{50-49,03}{0,99}\right)}{\Phi\left(\frac{50-49,03}{0,99}\right)} \right) = 48,7330$$

The proportion of theoretical censorship calculated as the Equation 5:

$$Pc = \Phi\left(\frac{50-49,03}{0,99}\right) = 0,843$$

Drawing control charts based on standardized control limits for the chart CEV $\overline{X}$ and S; (Figure 2 and Figure 3) these are 1.42 and 2.09, respectively.

*Table 1 (Data example)*

|    | 1    | 2    | 3    | 4    | 5    | X    | S   |
|----|------|------|------|------|------|------|-----|
| 1  | 50,0 | 50,0 | 50,0 | 50,0 | 50,0 | 50,0 | 0,0 |
| 2  | 50,0 | 50,0 | 50,0 | 50,0 | 50,0 | 50,0 | 0,0 |
| 3  | 50,0 | 50,0 | 50,0 | 50,0 | 50,0 | 50,0 | 0,0 |
| 4  | 50,3 | 50,0 | 50,0 | 50,0 | 50,0 | 50,1 | 0,2 |
| 5  | 50,0 | 50,2 | 50,0 | 50,7 | 50,0 | 50,2 | 0,3 |
| 6  | 50,4 | 50,0 | 50,0 | 50,0 | 50,0 | 50,1 | 0,2 |
| 7  | 50,0 | 50,3 | 50,8 | 50,0 | 50,0 | 50,2 | 0,3 |
| 8  | 50,6 | 50,0 | 50,0 | 50,0 | 51,2 | 50,4 | 0,5 |
| 9  | 50,0 | 50,5 | 50,9 | 50,8 | 50,6 | 50,5 | 0,4 |
| 10 | 50,0 | 50,0 | 50,0 | 50,0 | 50,7 | 50,1 | 0,3 |
| 11 | 50,0 | 50,4 | 50,0 | 50,0 | 50,0 | 50,1 | 0,2 |
| 12 | 50,0 | 50,0 | 50,0 | 50,0 | 50,0 | 50,0 | 0,0 |
| 13 | 50,0 | 50,0 | 50,0 | 50,0 | 50,0 | 50,0 | 0,0 |
| 14 | 50,0 | 50,0 | 50,0 | 50,0 | 50,0 | 50,0 | 0,0 |
| 15 | 50,0 | 50,0 | 50,0 | 50,0 | 50,0 | 50,0 | 0,0 |
| 16 | 50,0 | 50,0 | 50,0 | 50,0 | 50,0 | 50,0 | 0,0 |
| 17 | 50,9 | 50,0 | 50,0 | 50,0 | 50,0 | 50,2 | 0,4 |
| 18 | 50,0 | 50,0 | 50,0 | 50,0 | 50,0 | 50,0 | 0,0 |
| 19 | 50,0 | 50,0 | 50,0 | 50,0 | 50,0 | 50,0 | 0,0 |
| 20 | 50,0 | 50,0 | 50,0 | 50,0 | 50,0 | 50,0 | 0,0 |
| 21 | 50,3 | 50,0 | 50,0 | 50,0 | 50,0 | 50,1 | 0,1 |
| 22 | 50,0 | 50,0 | 50,0 | 50,0 | 50,5 | 50,1 | 0,2 |
| 23 | 50,0 | 50,0 | 50,0 | 50,2 | 50,0 | 50,0 | 0,1 |
| 24 | 50,0 | 50,0 | 50,0 | 51,0 | 50,0 | 50,2 | 0,4 |
| 25 | 50,0 | 50,0 | 50,0 | 50,0 | 50,0 | 50,0 | 0,0 |

Calculating the control limits for the control chart of the mean and standard deviation according to the Equation 15 and Equation 16 is obtained:

*Upper control limit for the chart CEV $\overline{x}$ = 1,42\*0,9915+49,0279=50,43583*
*Upper control limit for the chart CEV S = 2,09\*0,9915=2,0524*

Figure 5 and Figure 6 lists the results of the initial deployment, where points are not removed. In this case the points are within specifications. One can say that the data comes from a process under control. As a result, they may continue the monitoring process using the control limits given for the CEV model.

The lower control limit is unnecessary because no average subgroups of observations will be below Wc for graphic CEV X and for graphic CEV S. Thus, only increases were detected in the mean of the process, which in practice are usually more concerned.

Both Figure 5 and Figure 6, it is seen that there is no point outside the calculated control limits, so it can be said that the process is fully controlled.

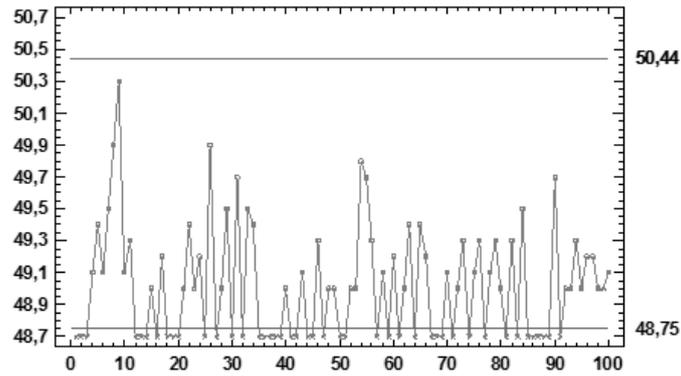
*Figure 5 (control chart CEV X for the model CEV)*

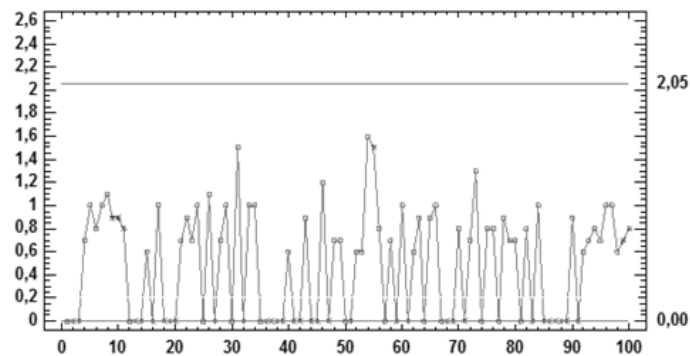
*Figure 6 (control chart CEV S for the model CEV)*

## 4 Conclusions

In this article, adapted control charting procedures to monitor the process mean and standard deviation applicable when observations are censored at fixed levels are proposed. The proposed charts are based on the idea of replacing all observations censored by their conditional expected values and then charting standard statistics of these CEV weights.

Situations in which the measuring equipment has a limited sensitivity are not entirely desirable, but are present and their treatment requires a series of precautions to avoid errors. The problem of estimating censored data is solved with maximum likelihood estimators and an iterative calculation process. This provides more accurate monitoring of the evaluation of the controlled variable with other alternatives achieved.

Control limits for CEV control charts given are derived from simulation of the sampling distributions of the subgroup statistics assuming that the in-control distribution is known and normal.

It is ideal that the percentage of censored data is not high. Highly censored data can generate significantly different estimates. In addition, given that the amount of information in each subgroup to detect changes in the process is small when the censorship is severe, the traditional Shewhart control graphic generates a large number of false alarms; for this case, the average CEV control chart is the appropriate one.

There are many other practical censorship schemes that should be investigated.

## Appendix

Below we present several tables 2, 3, 4, and 5, with the values of the coefficients for calculating the control limits with different sample sizes, and a probability of error type I (α) 0.0027.

*Table 2 (Coefficients for calculating the control limits, n=20)*

| n=20 | | |
|---|---|---|
| 1-%C | UCL Coefficient Mean | UCL Coefficient S |
| 0,02 | 0,69 | 1,50 |
| 0,03 | 0,68 | 1,49 |
| 0,04 | 0,68 | 1,49 |
| 0,07 | 0,68 | 1,49 |
| 0,10 | 0,68 | 1,49 |
| 0,16 | 0,68 | 1,49 |
| 0,24 | 0,68 | 1,49 |
| 0,31 | 0,68 | 1,49 |
| 0,50 | 0,68 | 1,49 |
| 0,69 | 0,68 | 1,49 |
| 0,84 | 0,68 | 1,49 |
| 0,98 | 0,68 | 1,49 |

*Table 3 (Coefficients for calculating the control limits, n=10)*

| n=10 | | |
|---|---|---|
| 1-%C | UCL Coefficient Mean | UCL Coefficient S |
| 0,02 | 1,02 | 1,80 |
| 0,03 | 0,97 | 1,71 |
| 0,04 | 0,97 | 1,71 |
| 0,07 | 0,97 | 1,71 |
| 0,10 | 0,97 | 1,71 |
| 0,16 | 0,98 | 1,72 |
| 0,24 | 0,97 | 1,72 |
| 0,31 | 0,97 | 1,72 |
| 0,50 | 0,97 | 1,71 |
| 0,69 | 0,97 | 1,72 |
| 0,84 | 0,97 | 1,71 |
| 0,98 | 0,97 | 1,71 |

*Table 4 (Coefficients for calculating the control limits, n=5)*

| n=5 | | |
|---|---|---|
| 1-%C | UCL Coefficient Mean | UCL Coefficient S |
| 0,02 | 1,61 | 2,36 |
| 0,03 | 1,47 | 2,15 |
| 0,04 | 1,42 | 2,09 |
| 0,07 | 1,42 | 2,08 |
| 0,10 | 1,42 | 2,07 |
| 0,16 | 1,42 | 2,09 |
| 0,24 | 1,43 | 2,09 |
| 0,31 | 1,42 | 2,09 |
| 0,50 | 1,43 | 2,09 |
| 0,69 | 1,42 | 2,09 |
| 0,84 | 1,43 | 2,09 |
| 0,98 | 1,42 | 2,08 |

*Table 5 (Coefficients for calculating the control limits, n=3)*

| n=3 | | |
|---|---|---|
| 1-%C | UCL Coefficient Mean | UCL Coefficient S |
| 0,02 | 2,46 | 3,23 |
| 0,03 | 2,11 | 2,78 |
| 0,04 | 1,94 | 2,54 |
| 0,07 | 1,92 | 2,53 |
| 0,10 | 1,94 | 2,55 |
| 0,16 | 1,95 | 2,56 |
| 0,24 | 1,95 | 2,56 |
| 0,31 | 1,95 | 2,56 |
| 0,50 | 1,95 | 2,56 |
| 0,69 | 1,95 | 2,56 |
| 0,84 | 1,95 | 2,57 |
| 0,98 | 1,95 | 2,56 |